\documentclass[conference]{IEEEtran}

\ifCLASSINFOpdf
\else
\fi

\usepackage{graphicx}
\usepackage{subfigure}
\usepackage{psfrag}
\usepackage{amsmath}
\usepackage{amsfonts}
\usepackage{amssymb}
\usepackage{float}

\newtheorem{definition}{Definition}

\newtheorem{lemma}{Lemma}

\makeatletter
\def\ifundefined{\@ifundefined}
\makeatother

\begin{document}

\title{Outage Capacity of Bursty Amplify-and-Forward with Incremental Relaying}

\author{\IEEEauthorblockN{T. Renk, H. J{\"a}kel, F.~K. Jondral}
\IEEEauthorblockA{Communications Engineering Lab\\
Karlsruhe Institute of Technology (KIT)\\
Email: tobias.renk@kit.edu}

%
%
%

\and

\IEEEauthorblockN{D. G{\"u}nd{\"u}z}
\IEEEauthorblockA{Centre Tecnol{\`o}gic de Tele- \\ comunicacions de Catalunya (CTTC)\\
Email: deniz.gunduz@cttc.es}

\and

\IEEEauthorblockN{A. Goldsmith}
\IEEEauthorblockA{Department of Electrical Engineering\\
Stanford University\\
Email: andrea@systems.stanford.edu}}

\def\snr{{\sf SNR}}
\def\dB{{\sf dB}}
\def\outcap{$\epsilon$-outage capacity }

\maketitle

\begin{abstract}
We derive the outage capacity of a bursty version of the amplify-and-forward (BAF) protocol for small signal-to-noise ratios when incremental relaying is used. We show that the ratio between the outage capacities of BAF and the cut-set bound is independent of the relay position and that BAF is outage optimal for certain conditions on the target rate $R$. This is in contrast to decode-and-forward with incremental relaying, where the relay location strongly determines the performance of the cooperative protocol. We further derive the outage capacity for a network consisting of an arbitrary number of relay nodes. In this case the relays transmit in subsequent partitions of the overall transmission block and the destination accumulates signal-to-noise ratio until it is able to decode. \newline

\textit{Keywords---} cooperative communications, incremental relaying, bursty amplify-and-forward, \outcap
\end{abstract}

\IEEEpeerreviewmaketitle

\section{Introduction}\label{sec:intro}

\PARstart{O}{ne}-bit feedback in a relay network, also called incremental relaying \cite{laneman04}, improves utilization of network degrees of freedom by minimizing the resources required for retransmission. Specifically, the one bit of feedback dictates whether the packet was received correctly, in which case retransmission is not needed, or if it was received incorrectly, making retransmission necessary. Analysis of this and more general forms of feedback is complicated by the fact that the average transmission rate is random, since it depends on the number of retransmissions required and, hence, the probability of successful source transmission.

The problem of variable rate has been addressed in \cite{laneman04} by introducing a long-term average rate $\bar{R}$. However,
average rate with asymptotically small error, i.e. Shannon capacity, is not a good metric for our analysis,
since we consider Rayleigh block fading where errors are inevitable at any nonzero transmission rate. Hence,
we allow outage events and derive expressions for the maximal transmission rate that achieves an outage probability lower than
a given target error rate $\epsilon$. This rate, called the $\epsilon$-outage capacity, was introduced in \cite{ozarow1994}. The \outcap of decode-and-forward with incremental relaying in the low SNR regime was derived in \cite{renk_vtc_fall_2009_1}. There, the authors solved the problem for the average \outcap by introducing a factor that accounts for the variability due to channel states. It is shown that the performance of decode-and-forward with incremental relaying strongly depends on the relay location. If the relay is located close to the source, this protocol is rate optimal. If it is located close to the destination, the ratio between the outage capacities of decode-and-forward with incremental relaying and the cut-set bound with feedback approaches $1/\sqrt{2}$ for the case of one relay. In \cite{avestimehr07} Avestimehr and Tse dealt with the outage capacities of the fading relay channel \textit{without} feedback. They showed that the `normal' version of amplify-and-forward is not applicable for low values of SNR since in this case the relay amplifies the noise,
which makes decoding at the destination more difficult. They then considered a bursty version of amplify-and-forward (BAF) and showed that this protocol is outage optimal for the frequency division duplex channel without feedback. This is in line with \cite{verdu2002}, where the author revises the fact that the capacity of an ideal bandlimited additive white Gaussian noise channel can be approached by pulse position modulation with a very low duty cycle in the low power regime (as stated in \cite{verdu2002}, this fact dates back to a publication by Golay in 1949 \cite{golay1949}). Finally, in \cite{atia2007} it is shown that BAF is also outage optimal for a wide class of independent channels where the distribution functions are smooth.\footnote{For more information on the smoothness of the distribution functions the interested reader is referred to \cite{atia2007}.}

The question addressed in this paper is the following: What is the \outcap of a BAF protocol \textit{with} incremental relaying when TDMA is applied, i.e., source and relay transmit in orthogonal time slots? For that purpose, we first consider the one-relay case. If the bursty
transmission from the source has not been successful, the relay transmits in the second time slot an amplified version of its own receive signal. The destination then accumulates the SNR values from the source and the relay transmission and tries to decode. If it is still not
able to decode, an outage is declared. For the general case of $K$ relays, the transmission procedure is as follows. After the source
transmission, the destination tries to decode. If it is not able to decode, the 'first' relay transmits an amplified version of its own receive
signal to the destination. (The question which relay should transmit as the 'first' relay can be solved, for instance, by taking
individual channel conditions into account.) As described before, the destination then tries to decode after having accumulated the two SNR values. If the destination is, however, still not able to decode, it sends a negative feedback and the `second' relay transmits. This procedure continues until the destination has either accumulated sufficient SNR to decode or all relays have transmitted. If the destination is then
still not able to decode, an outage is declared. However, once the destination is able to decode during the transmission procedure, it broadcasts a positive feedback indicating that the next time slot is reserved for the source in order to transmit the next packet.

The remainder of the paper is organized as follows. In Section \ref{sec:sys_mod} the system model is introduced and the relay protocol is described in detail. Section \ref{sec:out_cap_baf} deals with the derivation of the \outcap and points out the possible gains due to feedback. In Section \ref{sec:comp} the \outcap of BAF with incremental relaying is compared to the cut-set bound. Furthermore, in Section \ref{sec:ext} our work is extended to networks with an arbitrary number of relays and, finally, Section \ref{sec:conc} summarizes our finding and concludes the paper. Throughout the paper, we assume the feedback link to be perfect, i.e., source and relay receive information about success or failure of prior source transmission reliably. Investigations of imperfect feedback are done in \cite{renk_imp_fb}.

\section{System Model}\label{sec:sys_mod}
First we consider a network consisting of one source $\rm S$, one relay $\rm R$, and one destination $\rm D$. We use a block Rayleigh fading profile, i.e., the channel gains $h_i, \, i \in \{\rm sd, sr, rd\}$, are modeled as independent, zero-mean, circularly-symmetric complex Gaussian random variables that remain constant for the duration of one transmission block of length $T$. The variances $\sigma_i^2$ of the channel gains are proportional to $d_i^{-\alpha}$ with $d_i$ being the distance between two nodes and $\alpha$ denoting the path-loss exponent which typically lies between $3$ and $5$ for cellular mobile networks \cite{rappaport1996,goldsmith_05}. White Gaussian noise is added at each receiving node. Noise realizations are assumed to be independent and identically distributed (i.i.d.) and drawn from a zero-mean, circularly-symmetric Gaussian distribution with variance $N_0$. An average transmit power constraint of $P$ is used at the source and the relay over a transmission block, respectively, and SNR is defined as $\snr=P/N_0$. We further impose the half-duplex constraint on the relay, which means that the relay can either receive or transmit at any time instant, but cannot do both simultaneously.

\begin{figure}[t]
\centering
\psfrag{1 block}{one block}
\psfrag{AF:}{\textit{(a) AF with incremental relaying:}}
\psfrag{BAF:}{\textit{(b) BAF with incremental relaying:}}
\psfrag{S}{${\rm S}$}
\psfrag{R}{${\rm R}$}
\psfrag{or}{or}
\psfrag{P}{{$P$}}
\psfrag{T/2}{{$\frac{T}{2}$}}
\psfrag{Pa}{$\frac{P}{\tau}$}
\psfrag{aT/2}{$\frac{\tau T}{2}$}
\psfrag{fb}{${\rm FB}$}
\includegraphics[width=0.30\textwidth]{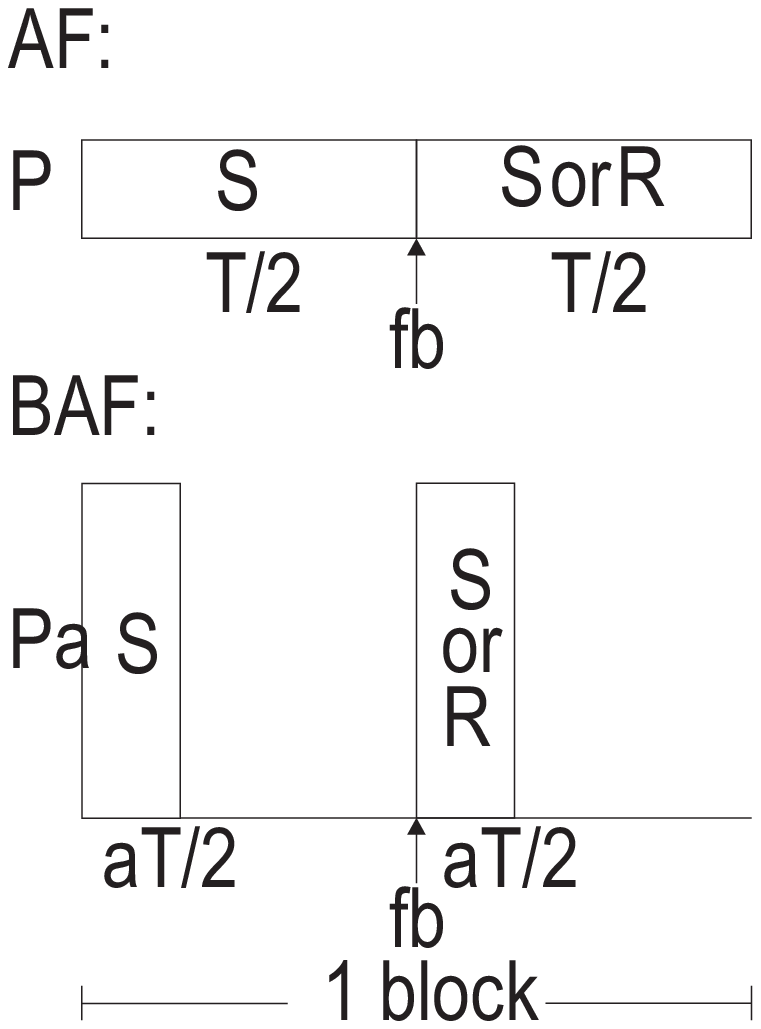}
\caption{Transmission model for incremental relaying. If the source-destination link is not in outage (feedback ${\rm FB}=1$), the source transmits during the second sub-block, too. If the source-destination link is in outage (feedback ${\rm FB}=0$), the relay aids communication during the second sub-block.}
\label{fig:blocks}
\end{figure}

The idea of AF and BAF with incremental relaying is shown in Fig.~\ref{fig:blocks}. The overall transmission block is divided into two time slots of equal length. During the first time slot the source broadcasts its message with power $P$ to the destination and the relay (subfigure (a)). The destination then sends a one-bit feedback ($\rm FB$) indicating success or failure of source transmission. Depending on the feedback either the source transmits its next message or the relay retransmits an amplified version of its own receive signal, i.e., an amplified version of the source's first message. As stated in Section \ref{sec:intro}, this protocol has poor performance for low values of SNR. Performance can be improved enormously if source and relay transmit bursts during their time slots, i.e., both transmit only for a fraction of $(\tau T)/2$ and with power $P/\tau$ ($\tau \rightarrow 0$) in order to meet the average power constraint (subfigure (b)). This is then comparable to pulse position modulation with a very low duty cycle (see \cite{verdu2002}).

\section{Outage Capacity of BAF}\label{sec:out_cap_baf}
In this section we derive the \outcap of BAF with incremental relaying. The way is similar to the one presented in \cite{renk_vtc_fall_2009_1}. We first derive an expression for the \outcap without feedback and then introduce a pre-factor that takes feedback into account. The instantaneous channel capacity for a half-duplex relay channel where BAF is applied is given by
\begin{eqnarray}
C_{\rm BAF}(\snr,\tau)&=&\frac{\tau}{2}\log_2\left(1+\frac{\snr}{\tau}\right. \nonumber \\
\hspace*{-3em}&&\hspace*{-3em}\left.\left(|h_{\rm sd}|^2+\frac{|h_{\rm rd}|^2|h_{\rm sr}|^2}{|h_{\rm rd}|^2+|h_{\rm sr}|^2+\tau/\snr}\right)\right).
\end{eqnarray}
This expression is similar to the ones given in \cite{avestimehr07,atia2007}. However, we consider an additional pre-log factor of $1/2$, which is due to the half-duplex constraint, and use the logarithm to the base $2$ in order to express capacity in bit/s/Hz. We set
\begin{equation}
\alpha(\mathbf{h},\tau) := |h_{\rm sd}|^2+\frac{|h_{\rm rd}|^2|h_{\rm sr}|^2}{|h_{\rm rd}|^2+|h_{\rm sr}|^2+\tau/\snr}
\end{equation}
and drop the dependence on $\mathbf{h}=(|h_{\rm sd}|^2,|h_{\rm sr}|^2,|h_{\rm rd}|^2)$ and $\tau$ in the following for the sake of description. An outage event occurs if $C_{\rm BAF}(\snr,\tau)$ is not large enough to serve a required target rate $R$. Hence, we have
\begin{equation*}
p_{\rm out}^{\rm (BAF)} = \Pr\left(\alpha < \frac{2^{2R/\tau}-1}{\snr/\tau}\right) .
\end{equation*}
Since we are interested in $\epsilon \rightarrow 0$, i.e., a target error rate that approaches zero in the low SNR regime ($\snr \rightarrow 0$), we have to ensure, by choosing $\tau$ in a suitable fashion, that the right hand side within the $\Pr(\cdot)$ expression tends to zero. A proper choice of $\tau$ would be $\tau=\sqrt{R \, \snr}$ (like in \cite{avestimehr07}). Plugging this into the above equation results in $\sqrt{R/\snr} \rightarrow 0$. Hence, outage probability can be expressed as
\begin{equation*}
p_{\rm out}^{\rm (BAF)} = \Pr\left(\alpha < g(R,\snr)\right)
\end{equation*}
with $g(R,\snr)$ being given by
\begin{equation*}
g(R,\snr) = \sqrt{\frac{R}{\snr}} \left( 2^{2\sqrt{R/\snr}} -1 \right).
\end{equation*}
In order to derive the $\epsilon$-outage capacity, we first state the following lemma.\footnote{Proof can be found in \cite{avestimehr07}.}
\begin{lemma}\label{lemma_cdf_general}
Let $U$, $V$, and $W$ be independent exponentially distributed random variables with mean $\sigma_u^2$, $\sigma_v^2$, and $\sigma_w^2$. If $g(x)$ is a continuous function at $x=0$ and $g(x) \rightarrow 0$ as $x \rightarrow 0$, then
\begin{equation}\label{eq:lemma_baf}
  \lim\limits_{x \rightarrow 0} \frac{1}{g(x)^2} \Pr\left(U+\frac{VW}{V+W+x}<g(x)\right) = \frac{\sigma_v^2+\sigma_w^2}{2\sigma_u^2\sigma_v^2\sigma_w^2}.
\end{equation}
\end{lemma}
We are now able to write
\begin{equation}\label{eq:lemma_baf}
  \lim\limits_{\substack{\epsilon \rightarrow 0 \\ \snr \rightarrow 0 \\ g(R,\snr) \rightarrow 0}} \frac{p_{\rm out}^{\rm (BAF)}}{g(R,\snr)^2} = \frac{\sigma_{\rm rd}^2+\sigma_{\rm sr}^2}{2\sigma_{\rm sd}^2\sigma_{\rm rd}^2\sigma_{\rm sr}^2}
\end{equation}
and the \outcap in bit/s/Hz of BAF \textit{without} incremental relaying after some proper manipulations becomes
\begin{equation}\label{eq:c_baf_wo_fb}
{\cal C}_{\epsilon}^{\rm (BAF)} \approx \frac{1}{2} \log_2\left( 1 + \snr \sqrt{\frac{2\sigma_{\rm sd}^2\sigma_{\rm rd}^2\sigma_{\rm sr}^2 \epsilon}{\sigma_{\rm rd}^2+\sigma_{\rm sr}^2}}\right) ,
\end{equation}
where we used the approximation\footnote{This approximation is related to the approximation $\ln (1+x) \approx x$ for small values of $x$.}
\begin{equation}\label{eq:approx}
\frac{x}{\log_2(e)} \approx \frac{1}{2} \log_2\left(1+x \right).
\end{equation}
As mentioned before, \eqref{eq:c_baf_wo_fb} does not consider the variability of the transmission rate in a long-term perspective. This variability is due to the feedback from the destination to the source and the relay. In order to take it into account, the average amount of transmitted sub-blocks required for sending one specific source message must be considered. If the source transmission during the first sub-block has been successful, only one sub-block is required independent of the relay. However, if the source transmission has failed, relay transmission over the second sub-block is necessary. If the destination is still not able to decode after the second sub-blocks, an outage will be declared. We define a random variable $N$ that describes the number of sub-block required for transmitting a specific message. The average of $N$ --for the one-relay case-- only depends on the source transmission during the first sub-block. We have $\mathbb{E}(N)=1+\Pr(\text{``$\rm S$ to $\rm D$ fails''})$. With these considerations, we are able to express the \outcap of BAF \textit{with} incremental relaying as
\begin{eqnarray}
{\cal C}_{\epsilon,{\rm IR}}^{\rm (BAF)} &=& \frac{2}{\mathbb{E}(N)} {\cal C}_{\epsilon}^{\rm (BAF)} \label{eq:c_baf_ir_1} \\
&\approx& \frac{1}{\mathbb{E}(N)} \log_2\left( 1 + \snr \sqrt{\frac{2\sigma_{\rm sd}^2\sigma_{\rm rd}^2\sigma_{\rm sr}^2 \epsilon}{\sigma_{\rm rd}^2+\sigma_{\rm sr}^2}}\right).
\end{eqnarray}
The factor $2/\mathbb{E}(N)$ in \eqref{eq:c_baf_ir_1} is due to possible savings in the required amount of sub-blocks for transmitting a specific source message. If only transmission over one sub-block is required, i.e., source transmission has been successful, a gain of $2$ can be achieved ($\mathbb{E}(N)=1$), since then the source can transmit its next message after reception of the positive feedback from the destination (${\rm FB} = 1$)\footnote{Recall that we assume block fading.}. If both sub-blocks are necessary for transmitting one and the same message, i.e., source transmission has failed and the relay aids communication (${\rm FB} = 0$), we perform at least as good as a BAF protocol without feedback ($\mathbb{E}(N)=2$). Moreover, it can easily be verified that if we consider a one-dimensional geometry, where the relay is placed on a straight line between source and destination, and the path-loss model presented in Section \ref{sec:sys_mod}, the optimal relay location that maximizes the \outcap is $d_{\rm sr}^{\ast}=0.5$ independent of the path-loss factor $\alpha$.

\section{Comparison}\label{sec:comp}

In order to compare the \outcap of BAF with incremental relaying to the cut-set bound (CSB), we define the following performance criterion (cf. \cite{renk_vtc_fall_2009_1,renk_ew_2010}).
\begin{definition}\label{def:delta_epsi}
The ratio between incremental relaying and the cut-set bound for the same target outage probability $\epsilon$ is defined as
\begin{equation}
\Delta(\epsilon):=\frac{{\cal C}_{\epsilon,{\rm IR}}^{\rm (BAF)}}{{\cal C}_{\epsilon}^{\rm (CSB)}} \leq 1.
\end{equation}
\end{definition}
The outage capacity of the CSB with incremental relaying is given by \cite{renk_vtc_fall_2009_1}
\begin{equation}\label{eq:c_e_csb}
{\cal C}_{\epsilon}^{\rm (CSB)} \geq \frac{1}{1+\epsilon} \log_2\left( 1+\snr \sqrt{\frac{2\sigma_{\rm sd}^2\sigma_{\rm sr}^2\sigma_{\rm rd}^2 \epsilon}{\sigma_{\rm rd}^2+\sigma_{\rm sr}^2}}\right),
\end{equation}
where we have applied $\mathbb{E}(N) \geq 1 + \epsilon$. This lower bound on the average amount of sub-block transmissions makes sense for the following reason. Our target error rate is given by $\epsilon$. Consequently, the outage probability for the source transmission in the first sub-block must be higher than or equal to $\epsilon$. In order to get a tighter bound for the \outcap of the CSB with incremental relaying, it is thus reasonable to use $\mathbb{E}(N) \geq 1 + \epsilon$. Comparison of BAF to the CSB leads to the ratio
\begin{equation}
\Delta(\epsilon) \leq \frac{1+\epsilon}{\mathbb{E}(N)} = \frac{1+\epsilon}{1+\Pr(\text{``$\rm S$ to $\rm D$ fails''})}.
\end{equation}
The outage probability of source transmission in the low SNR regime can easily be derived. We get
\begin{eqnarray}
\Pr(\text{``$\rm S$ to $\rm D$ fails''}) \hspace*{-0.7em}&=&\hspace*{-0.7em} \Pr\left(\frac{\tau}{2}\log_2\left(1+|h_{\rm sd}|^2\frac{\snr}{\tau}\right)<R\right) \nonumber \\
&\approx& \frac{\log_2(e) \, R}{\sigma_{\rm sd}^2 \snr} \nonumber,
\end{eqnarray}
where we again set $\tau = \sqrt{R \, \snr}$, let $\sqrt{R/\snr} \rightarrow 0$, and used the approximation given in \eqref{eq:approx}. Since $\Pr(\text{``$\rm S$ to $\rm D$ fails''})$ must be higher than $\epsilon$, we get an upper bound on the target error rate of $\epsilon \leq \frac{\log_2(e) \, R}{\sigma_{\rm sd}^2 \snr}$.

\begin{figure}[t]
\centering
\psfrag{Delta epsilon}{$\Delta(\epsilon)$}
\psfrag{0.2}{$0.2$}
\psfrag{0.4}{$0.4$}
\psfrag{0.6}{$0.6$}
\psfrag{0.8}{$0.8$}
\psfrag{1}{$1$}
\psfrag{-10}{$-10$}
\psfrag{-8}{$-8$}
\psfrag{-6}{$-6$}
\psfrag{-4}{$-4$}
\psfrag{-2}{$-2$}
\psfrag{0}{$0$}
\psfrag{2}{$2$}
\psfrag{4}{$4$}
\psfrag{6}{$6$}
\psfrag{8}{$8$}
\psfrag{10}{$10$}
\psfrag{SNR [dB]}{$\snr$ [dB]}
\psfrag{R=0.009}{$R=0.009$}
\psfrag{R=0.05}{$R=0.05$}
\psfrag{R=0.1}{$R=0.1$}
\includegraphics[width=0.5\textwidth]{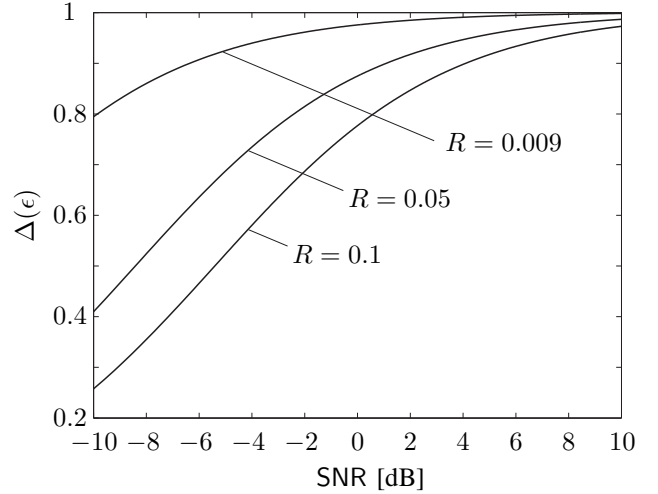}
\caption{The ratio $\Delta(\epsilon)$ versus SNR for $\epsilon=0.001$ and different values of rate $R$ in bit/s/Hz. The distance
source-destination has been normalized to $1$.}
\label{fig:Delta_epsilon_vs_snr_dB}
\end{figure}

Fig.~\ref{fig:Delta_epsilon_vs_snr_dB} depicts the ratio $\Delta(\epsilon)$ versus SNR in dB for $\epsilon=0.001$. The distance source-destination has been normalized to $1$, i.e., $\sigma_{\rm sd}^2=1$. Obviously, $\Delta(\epsilon)$ is a monotonically increasing function in $\snr$. We see that the values of $\Delta(\epsilon)$ for a given $\snr$ are lower if the rate $R$ is increased.

\section{Extension to $K$ Relays}\label{sec:ext}

\begin{figure}[t]
\centering
\psfrag{1 block}{one block}
\psfrag{S}{${\rm S}$}
\psfrag{R1}{${\rm R}_1$}
\psfrag{Rm}{${\rm R}_m$}
\psfrag{Rn}{${\rm R}_n$}
\psfrag{or}{or}
\psfrag{Pa}{$\frac{P}{\tau}$}
\psfrag{aT/K+1}{$\frac{\tau T}{K+1}$}
\psfrag{fb}{${\rm FB}$}
\includegraphics[width=0.48\textwidth]{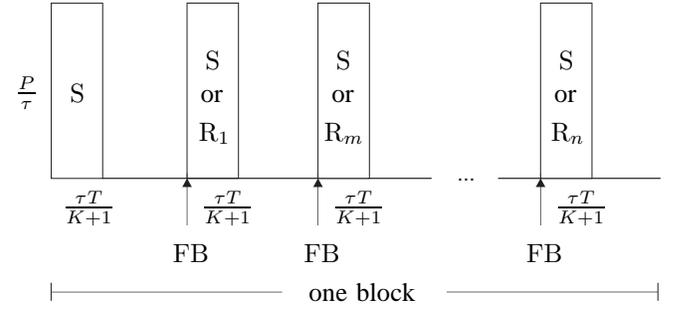}
\caption{Transmission model for BAF with incremental relaying and an arbitrary number of relay nodes. ${\rm R}_m$, $m \in \{1,2\}$, and ${\rm
R}_n$, $n \in \{1,2,\ldots,K\}$, describe the transmitting relay depending on prior transmissions.}
\label{fig:block3}
\end{figure}

In this section we extend the previous work to relay networks with an arbitrary number of relay nodes. The basic transmission model is illustrated in Fig.~\ref{fig:block3}. The basic idea is that the destination sends negative feedbacks (${\rm FB}=0$) until it has accumulated sufficient SNR to decode. For instance, in the second sub-block either the source $\rm S$ or the first relay ${\rm R}_1$ transmits depending on whether source transmission has been successful during the first sub-block or not. In the third sub-block either the source $\rm S$ or the first relay ${\rm R}_1$ or the second relay ${\rm R}_2$ transmits depending on whether the previously accumulated SNR has been sufficient for the destination to decode and so on. Once the destination has accumulated enough SNR to decode reliably, it sends a positive feedback (${\rm FB}=1$) indicating that no more relay transmissions are required. When this happens, the source starts transmitting its new message in the next sub-block. We can immediately conclude that such a procedure would lead to a maximal gain of $K+1$ (compared to a BAF protocol without incremental relaying) if source transmission in the first sub-block is successful. If all relays have to transmit, the possible gain reduces to $1$. If the SNR at the destination still is not sufficient to decode after the $K$-th relay has transmitted, an outage is declared.

The instantaneous channel capacity for BAF with $K$ relays and a TDMA transmission scheme, where one transmission block is divided into $K+1$ sub-blocks of equal length, is given by
\begin{equation}
C_{{\rm BAF},K}(\snr,\tau)=\frac{\tau}{K+1}\log_2\left(1+\frac{\snr}{\tau} \alpha_K\right),
\end{equation}
where we used the abbreviation
\begin{equation*}
\alpha_K := |h_{\rm sd}|^2+\sum\limits_{k=1}^{K}\frac{|h_{{\rm r}_k{\rm d}}|^2|h_{{\rm sr}_k}|^2}{|h_{{\rm r}_k{\rm d}}|^2+|h_{{\rm sr}_k}|^2+\tau/\snr}
\end{equation*}
and again dropped the dependence on the $(K+1)$-tuple $\mathbf{h}_K=(|h_{\rm sd}|^2,|h_{{\rm sr}_k}|^2,|h_{{\rm r}_k{\rm d}}|^2)$, $k=1,\ldots,K$, and $\tau$ for the sake of description. For the low SNR regime, the capacity can be approximated by
\begin{equation}
C_{{\rm BAF},K}(\snr,\tau) \approx \frac{\snr}{K+1} \log_2(e) \, \alpha_K
\end{equation}
and the outage probability eventually becomes
\begin{equation}
p_{{\rm out},K}^{\rm (BAF)} \approx \Pr\left(\alpha_K < g_K(R,\snr)\right),
\end{equation}
where we used $g_K(R,\snr) = (K+1)R/(\log_2(e) \, \snr)$. Since the solution gets involved due to the structure of $\alpha_K$, we use the following inequality to upper bound it:
\begin{equation*}
\min\{x,y\} \geq \frac{xy}{x+y+\delta}, \quad x,y \in \mathbb{R}^+,
\end{equation*}
where $\delta$ is an arbitrarily small and positive number.
By defining
\begin{equation*}
\alpha_K' := |h_{\rm sd}|^2+\sum_{k=1}^{K}\min\{|h_{{\rm r}_k{\rm d}}|^2,|h_{{\rm sr}_k}|^2\},
\end{equation*}
we get 
\begin{equation}
p_{{\rm out},K}^{\rm (BAF)} \geq \Pr\left(\alpha_K' < g_K(R,\snr)\right).
\end{equation}
Using results given in \cite{avestimehr07} and applying \eqref{eq:approx} again finally yields
\begin{equation*}\label{eq:lemma_baf_K}
  \lim\limits_{\substack{\epsilon \rightarrow 0 \\ \snr \rightarrow 0 \\ g_K(R,\snr) \rightarrow 0}} \frac{p_{{\rm out},K}^{\rm (BAF)}}{g_K(R,\snr)^{K+1}} \geq \frac{\prod\limits_{k=1}^{K}(\sigma_{{\rm r}_k{\rm d}}^2+\sigma_{{\rm sr}_k}^2)}{(K+1)!\sigma_{\rm sd}^2\prod\limits_{k=1}^{K}\sigma_{{\rm r}_k{\rm d}}^2 \sigma_{{\rm sr}_k}^2}
\end{equation*}
and, therefore, an \outcap of BAF \textit{without} incremental relaying of
\begin{eqnarray}
{\cal C}_{\epsilon,K}^{\rm (BAF)} &\leq& \frac{1}{K+1}\log_2 \Bigg( 1+\snr  \nonumber \\
\hspace*{-1em}&&\hspace*{-1em} \sqrt[K+1]{\frac{(K+1)!\sigma_{\rm sd}^2\prod_{k=1}^{K}\sigma_{{\rm r}_k{\rm d}}^2 \sigma_{{\rm sr}_k}^2 \epsilon}{\prod_{k=1}^{K}(\sigma_{{\rm r}_k{\rm d}}^2+\sigma_{{\rm sr}_k}^2)}}\Bigg).
\end{eqnarray}
The \outcap \textit{with} incremental relaying then is
\vspace*{-0.3em}
\begin{eqnarray}
{\cal C}_{\epsilon,{\rm IR},K}^{\rm (BAF)} &\leq& \frac{1}{\mathbb{E}_K(N)} \log_2 \Bigg( 1 + \snr \nonumber \\
\hspace*{-1em}&&\hspace*{-1em}\sqrt[K+1]{\frac{(K+1)!\sigma_{\rm sd}^2\prod_{k=1}^{K}\sigma_{{\rm r}_k{\rm d}}^2 \sigma_{{\rm sr}_k}^2 \epsilon}{\prod_{k=1}^{K}(\sigma_{{\rm r}_k{\rm d}}^2+\sigma_{{\rm sr}_k}^2)}}\Bigg).
\end{eqnarray}
In the above equation, $\mathbb{E}_K(N)$ denotes the average amount of required sub-blocks in order to send a specific source message to the destination. It is given by
\vspace*{-0.3em}
\begin{eqnarray}
\mathbb{E}_K(N) &=& 1 + \sum\limits_{k=1}^{K}\Pr\bigg(|h_{\rm sd}|^2 \nonumber \\
\hspace*{-2em}&&\hspace*{-2em}+\sum\limits_{l=1}^{k-1}\frac{|h_{{\rm r}_l{\rm d}}|^2|h_{{\rm sr}_l}|^2}{|h_{{\rm r}_l{\rm d}}|^2+|h_{{\rm sr}_l}|^2+\tau/\snr} < g_K(R,\snr)\bigg). \nonumber
\end{eqnarray}
This practically means that the destination accumulates SNR until it is able to decode the intended source message. If the destination is still not able to decode after the $K$-th relay has transmitted, an outage event occurs. 
The \outcap of the CSB with incremental relaying can readily be shown to be upper bounded by (cf. \cite{renk_vtc_fall_2009_1})
\vspace*{-0.5em}
\begin{eqnarray}
{\cal C}_{\epsilon,K}^{\rm (CSB)} &\leq& \frac{1}{1+K\epsilon} \log_2 \Bigg( 1 + \snr \nonumber \\ &&\sqrt[K+1]{\frac{(K+1)!\sigma_{\rm sd}^2\prod_{k=1}^{K}\sigma_{{\rm r}_k{\rm d}}^2 \sigma_{{\rm sr}_k}^2 \epsilon}{\prod_{k=1}^{K}(\sigma_{{\rm r}_k{\rm d}}^2+\sigma_{{\rm sr}_k}^2)}}\Bigg),
\end{eqnarray}
where we only considered the broadcast and the multiple access cut (normally one would have $2^K$ cuts). We now apply Definition \ref{def:delta_epsi} in order to compare the performance of BAF with incremental relaying to the CSB with incremental relaying. Accordingly,
\begin{equation*}
\Delta_K(\epsilon) \leq \frac{1+K\epsilon}{\mathbb{E}_K(N)}.
\end{equation*}
In contrast to the one-relay case, we see that the ratio between the $\epsilon$-outage capacities depends on the relay locations, which determine the average amount of required sub-blocks (i.e., $\mathbb{E}_K(N)$). Clearly, $\epsilon \leq \frac{1}{K} \left( \mathbb{E}_K(N) - 1 \right)$.


\section{Conclusions and Outlook}\label{sec:conc}
We presented the \outcap of a bursty version of the amplify-and-forward protocol when incremental relaying is used, i.e., there is a one-bit feedback from the destination indicating success or failure of source transmission. We compared this protocol to the cut-set bound with incremental relaying and were able to show that for the one-relay case the ratio between the $\epsilon$-outage capacities is independent of the relay location and that BAF is also outage rate optimal in a setting with feedback if proper target rate adaptation is applied. Furthermore, we extended our results to networks with an arbitrary number of relays where the destination indicates after each sub-block if it has been able to decode (i.e., the number of required relays is not fixed, but adapted dynamically to the channel conditions). The one-bit feedback in our investigations has been considered to be received perfectly by the source and the relay. Current research deals with the effects of imperfect feedback. Initial results can be found in \cite{renk_imp_fb}.
%




\end{document}